\documentclass[twocolumn,english,aps,prl,superscriptaddress]{revtex4}
\usepackage{amsmath}
\usepackage{amssymb}
\usepackage{graphicx}
\usepackage{esint}
\usepackage{xcolor}
\usepackage{mathtools}
\usepackage[colorlinks=true]{hyperref}
\usepackage{ulem}
\hypersetup{
    bookmarks=true,         
    unicode=false,          
    pdftoolbar=true,        
    pdfmenubar=true,        
    pdffitwindow=false,     
    pdfstartview={FitH},    
    pdftitle={S},    
    pdfauthor={S. Sachdev},     
    pdfsubject={},   
    pdfcreator={},   
    pdfproducer={}, 
    pdfkeywords={} {} {}, 
    pdfnewwindow=true,      
    colorlinks=true,       
    linkcolor=magenta, 
    citecolor=blue,        
    filecolor=magenta,      
    urlcolor=blue           
}

\makeatletter

\usepackage{babel}

\usepackage{hyperref}

\makeatother

\usepackage{babel}
\begin{document}

\title{Quasiparticle metamorphosis in the random $t$-$J$ model}

\author{Aman Kumar}

\affiliation{Department of Theoretical Physics, Tata Institute of Fundamental
Research, Homi Bhabha Road, Navy Nagar, Mumbai 400005, India}

\author{Subir Sachdev}

\affiliation{Department of Physics, Harvard University, Cambridge MA-02138, USA}
\affiliation{School of Natural Sciences, Institute for Advanced Study, Princeton, NJ-08540, USA}

\author{Vikram Tripathi}

\affiliation{Department of Theoretical Physics, Tata Institute of Fundamental
Research, Homi Bhabha Road, Navy Nagar, Mumbai 400005, India}

\date{\today}
\begin{abstract}
Motivated by the pseudogap-Fermi liquid transition in doped Mott insulators, we examine the excitations of a $t$-$J$ model with random and all-to-all hopping and exchange.
The stability of quasiparticles such as spin-$1/2$ fermions, spin-$1$ magnons, and emergent Jordan-Wigner (JW) spinless fermions is cast as a problem of localization in the many-body Hilbert space, which is studied by the FEAST eigensolver algorithm. At low dopings, magnons and JW fermions are stable and better defined than spin-$1/2$ fermions, which are unstable. Upon crossing a critical value of doping around $p_c=1/3$, their stabilities are interchanged. Near the critical doping, these quasiparticles are all found to be ill-defined. The critical point is thus associated with a localization transition in the many-body Hilbert space.
\end{abstract}
\maketitle

Doped Mott insulators are central to understanding the unusual normal state of high-temperature superconductors, particularly the cuprates. The low hole doping pseudogap phase of the cuprates is characterised by collinear antiferromagnetic order, low carrier concentration with poor metallic conduction, a small Fermi surface \cite{frachet2020hidden}, and an anomalously large thermal Hall conductivity even in the absence of doping \cite{grissonnanche2019giant}. At large hole doping, the system is a paramagnetic Landau Fermi liquid with a large Fermi surface and spinful fermionic quasiparticles. In the presence of magnetic order, spin-$1$ magnons are natural excitations \cite{arrachea2002infinite}, but the collinear magnetic order and the low hole concentration in the underdoped phase make it unlikely that magnons or holes are the mechanism for the observed large thermal Hall conductivity
\cite{grissonnanche2019giant,grissonnanche2020chiral,samajdar2019enhanced,dalla2015fractional}. 
Instead, in analogy with the Kitaev model \cite{kitaev2006anyons} where emergent free (Jordan-Wigner) Majorana fermions have been predicted to result in a quantized thermal Hall effect \cite{yokoi2021half,kasahara2018majorana}, and reportedly confirmed in a Mott insulating Kitaev material $\alpha$-RuCl$_3,$ it has been suggested that the large thermal Hall effect in the cuprates \cite{grissonnanche2020chiral} could similarly arise from phonons coupling to some emergent fermionic excitations in the pseudogap phase, including at zero doping. At a certain critical value of doping, magnons are not well-defined since magnetic order vanishes before this doping is reached \cite{frachet2020hidden}, and neither are Landau quasiparticles since transport properties \cite{varma2020colloquium,cha2020linear,guo2020linear} are inconsistent with a Landau Fermi liquid. These observations motivate us to pose the question of stability of different quasiparticles across doping regimes of a Mott insulator. We identify quasiparticle metamorphosis upon varying the doping with many-body localization transitions in the Hilbert space. 

We shall study the fully-connected and random $t$-$J$ model, which has all-to-all hopping $t_{ij}$ and exchange $J_{ij}$ realized as independent random variables with zero mean. At zero doping, the model with SU($M$) symmetry, and $M$ large, realizes a Sachdev-Ye-Kitaev model of fractionalized fermionic spinons \cite{SY,kitaev_talk}. We are interested here in the case with SU(2) symmetry and non-zero doping: this has emerged a useful theoretical paradigm for studying the normal state properties of doped Mott insulators \cite{tJreview}. 
Recent analytical treatments of this model \cite{joshi2020deconfined,tarnopolsky2020metal} present a picture of a spin glass phase in low doping regime (with bosonic spinon and fermionic holon quasiparticles), surviving  up to a critical value of doping $p_c$, separated from a disordered Fermi liquid  at  large doping. However whether any of these quasiparticles or something different is responsible for the anomalously large thermal Hall effect in the underdoped regime remains an open question. In this paper, we study the stability of three different quasiparticles across doping regimes in the random $t$-$J$ model of a doped Mott insulator by casting the problem as one of localization in the many-body Hilbert space \cite{altshuler1997,aman_kitaev}. We use a numerical exact diagonalization method based on the recently developed FEAST algorithm \cite{FEAST} that allows calculation of large numbers of eigenstates in user-specified energy windows required for this approach.

The Hamiltonian of the random $t$-$J$ model is
\begin{equation}{\label{eq:1}}
     H=\sum_{i\neq j;\alpha=\uparrow,\downarrow}^{N}{\frac{t_{ij}}{\sqrt{N}}Pc^\dagger_{i\alpha}c_{j\alpha}P}+\sum_{i<j}^{N}{\frac{J_{ij}}{\sqrt{N}}\mathbf{S}_{i}\cdot\mathbf{S}_{j}},    
\end{equation}
where $t_{ij}$ and $J_{ij}$ are both randomly chosen from a Gaussian distribution with mean zero and unit variance. $P$ is the projector prohibiting double occupancy at any site, and $N$ is the total number of sites. A fraction $p$ of the sites is empty, corresponding to hole doping. The spin operator at each site is 
$\mathbf{S}_i=\frac{1}{2}\sum_{\alpha\beta}c^\dagger_{i\alpha}\boldsymbol{\sigma}_{\alpha \beta}c_{i\beta}.$ 

At low hole doping, the system has magnetic order and we show that spin-$1$ magnon excitations are good quasiparticles - a fact confirmed in numerous numerical \cite{arrachea2002infinite,shackleton2021quantum,Dumi21} and theoretical studies \cite{joshi2020deconfined,tarnopolsky2020metal}. Remarkably, in this doping regime, we also find that emergent Jordan-Wigner (JW) fermionic excitations are good quasiparticles reminiscent of the Mott insulating Kitaev materials \cite{banerjee2016proximate}. In contrast, spinful fermions are poor quasiparticles. In the opposite limit of large hole doping, we confirm that spinful fermionic quasiparticles are stable while the magnons and JW fermions are not. 

Our study indicates, similar to the cuprates, a critical point separating these regimes, which we identify as a many-body localization transition. We find that all the above quasiparticles are bad at criticality.

We formulate the problem of quasiparticle stability as one of many-body localization (MBL) \cite{altshuler1997,aman_kitaev} in the Hilbert space. Quasiparticles are approximations to the exact many-body eigenstates. Their lifetime and decay can be understood as follows. Any quasiparticle state $|\phi_i(t)\rangle$ can always be expanded as a linear superposition of the exact many-body stationary states $|\psi_j\rangle$ of the Hamiltonian, $|\phi_i(0)\rangle = \sum_{j}b_{ij}|\psi_j\rangle.$ At a later time $t,$ the quasiparticle wavefunction is $|\phi_i(t)\rangle = \sum_{j}\exp(-i E_j t)b_{ij}|\psi_j\rangle,$ where $E_j$ are the energy eigenvalues. If the weights $b_{ij}$ are significant only in a finite window of energies, $\Delta E,$ then the wavepacket decays in a time 
$\Delta t = 1/\Delta E.$ Since the number of energy eigenstates of the interacting system generally grows exponentially with the system size while the bandwidth grows much more slowly, the energy spread of the quasiparticle will be vanishingly small in the thermodynamic limit if the number of significant $b_{ij}$ (in other words, the support size of the quasiparticle in the many-body Hilbert space) does not scale in proportion to the dimension of the Hilbert space. On the other hand, if the support size scales in proportion to the Hilbert space dimension, the quasiparticle has a finite lifetime in the thermodynamic limit. These ideas are put on a quantitative footing below.

The quasiparticle stability is determined as follows. We first define our magnon and fermionic quasiparticles. We construct the set of $N$ magnon states by applying the spin raising operator at each site on the ground state $|\chi\rangle$ of only the \textit{exchange} part of the Hamiltonian, $\{c^\dagger _{\uparrow,1} c_{\downarrow,1}|\chi\rangle,....,c^\dagger _{\uparrow,N} c_{\downarrow,N}|\chi\rangle\}.$
We then use the Gram-Schmidt orthonormalization to construct a set of $N$ orthonormal states. We project the Hamiltonian in Eq. \ref{eq:1} to this magnon basis, and diagonalization yields the magnon quasiparticle states. We represent this set of $N$ orthonormal magnon states as $\{|\chi\prime_i\rangle\}$. To measure the resemblance of a given magnon state $|\chi\prime_i\rangle$ with the exact many-body states $|\psi_j\rangle$ of the full Hamiltonian, we express it as a linear superposition,
\begin{equation}
|\chi\prime_i\rangle=\sum_{j=1}^{D}{a_{ij}|\psi_j\rangle},
\end{equation}
where $D$ is the dimension of the Fock space of physically relevant states (i.e. do not include doubly occupied sites). The JW fermion creation (annihilation) operators are constructed by taking the product of the spin raising (lowering) operator at some site $i$ and a (nonlocal) 1D Jorgan-Wigner string $\prod_{j<i}[p_j I + n_j \sigma^{z}_{j}],$ where $n_j = 1-p_j.$  
Note our Jordan Wigner strings differ from the string operator introduced in ref \cite{grusdt2019microscopic} for the study of spinon chargon correlations in the antiferromagnetic $t$-$J$ model. 
We repeat a similar procedure for the fermionic quasiparticles. In particular, the hole states are obtained by applying annihilation operators ($c_{i\alpha}$) on the ground state of only the \textit{hopping} part of the Hamiltonian. There are $2N$ number of hole excitations. After the aforementioned Gram-Schmidt orthonormalization, projection of the full Hamiltonian to this basis, and subsequent diagonalization, we obtain a set of $2N$ fermionic hole quasiparticle states.  The hole quasiparticles are then expressed as a linear superposition of the exact many-body states.
Note that the magnon or hole quasiparticles are not eigenstates of the random $t$-$J$ model.

\begin{figure*}
	\includegraphics[width=1\columnwidth]{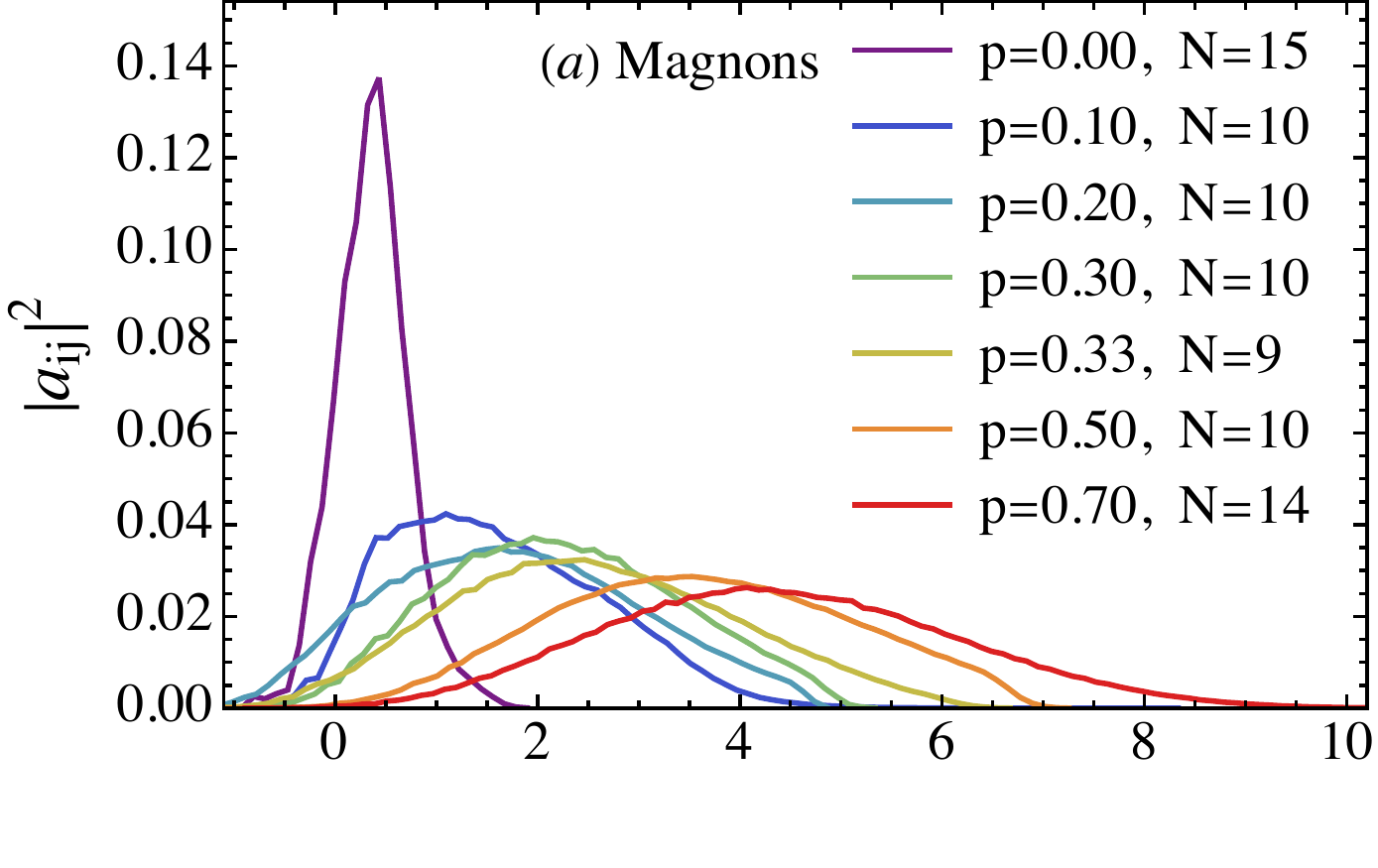}
	\includegraphics[width=\columnwidth]{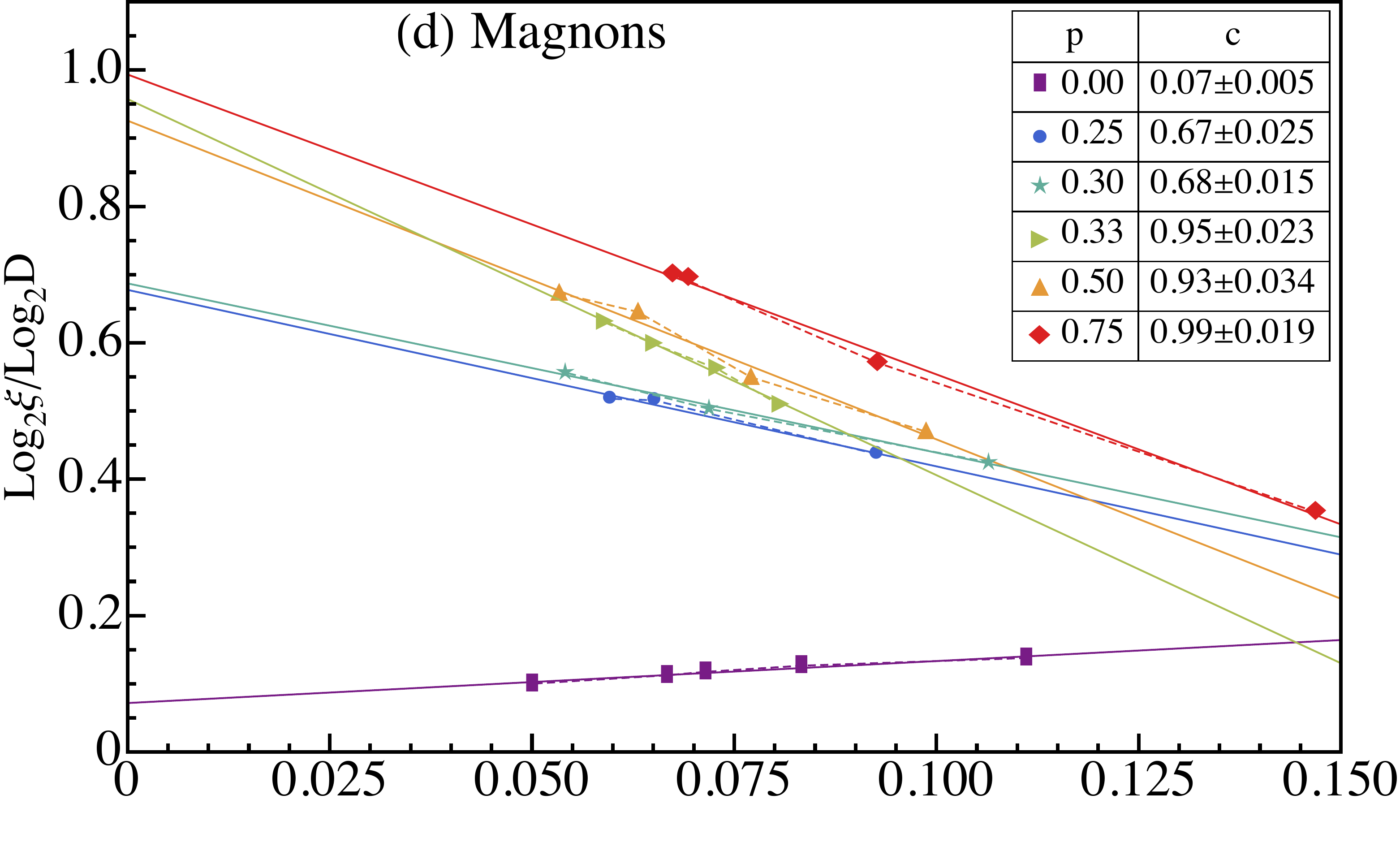}
	\includegraphics[width=1\columnwidth]{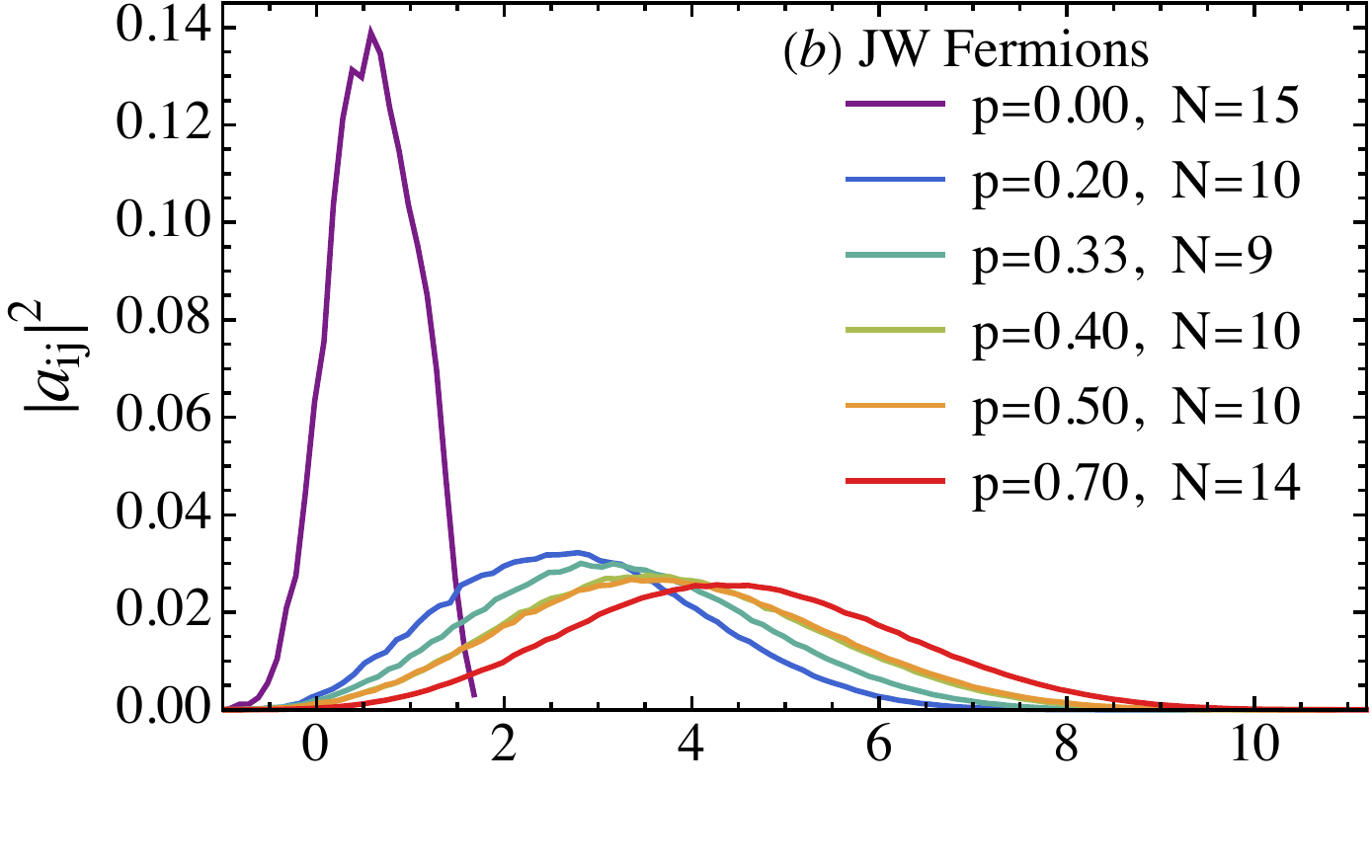}
		\includegraphics[width=\columnwidth]{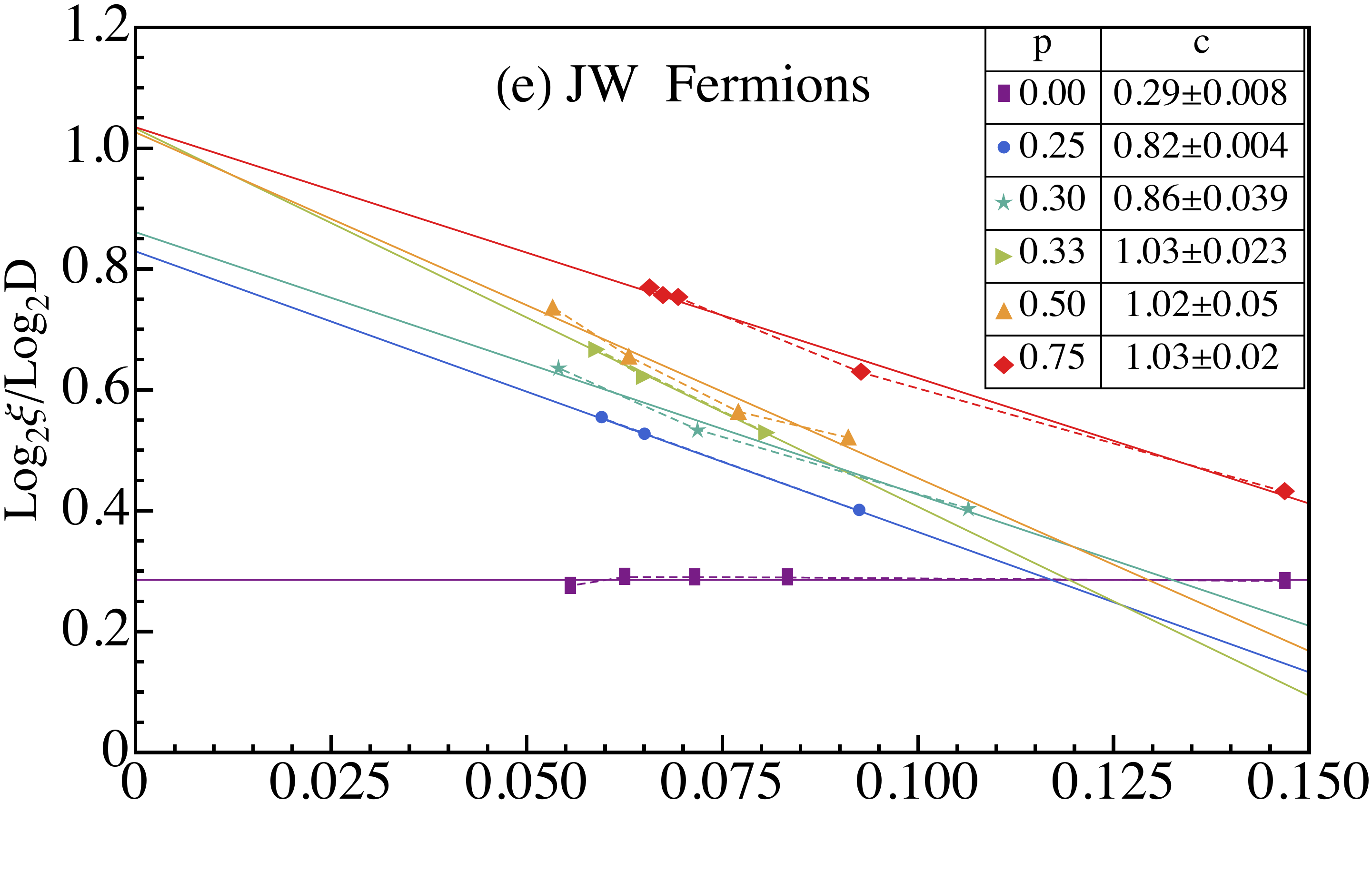}
	\includegraphics[width=\columnwidth]{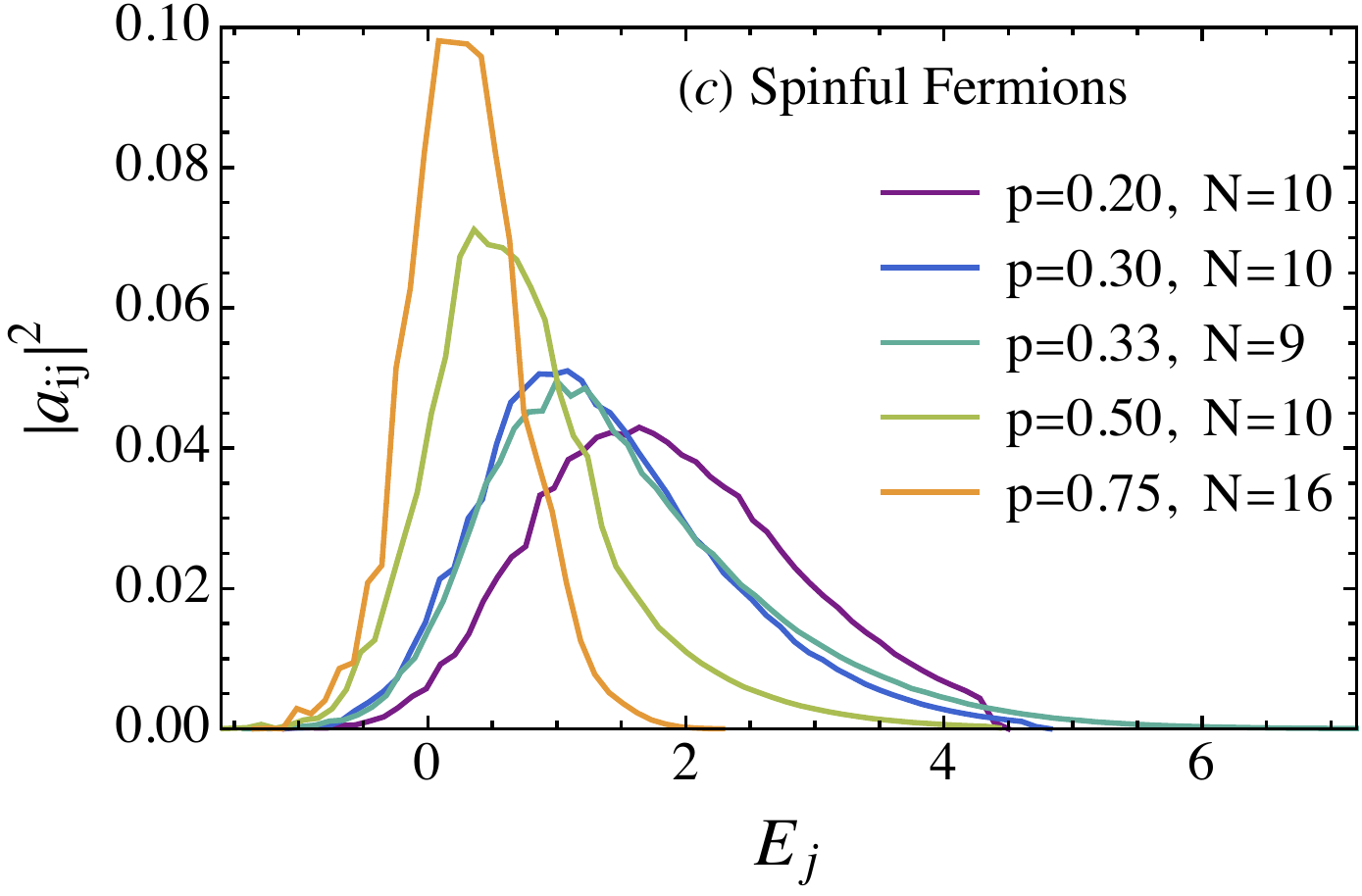}
		\includegraphics[width=\columnwidth]{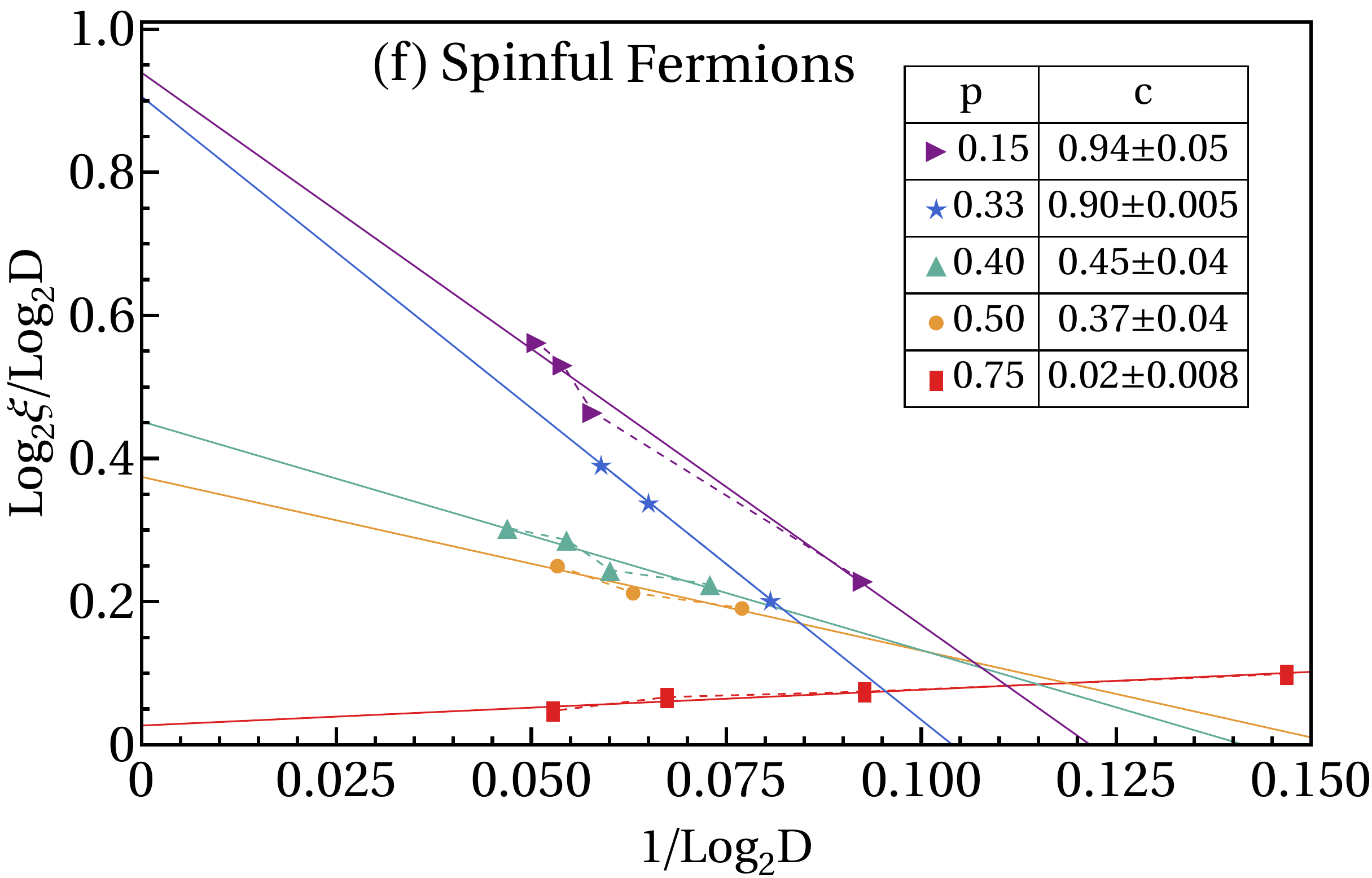}
	\caption{\label{fig:overlap}Plot showing (disorder averaged) squares of the overlap of a magnon (a), a JW fermion (b), and a hole (c) quasiparticle state (labeled $i$) with exact many-body states (excitation energy $E_j$ measured from disorder averaged ground state energy of many body state) of the random $t-J$ model for different values of doping density $p,$ and system size $N.$ For low doping density $p<0.33,$ (the spin glass phase) magnons (a) and JW fermions have small support sizes (sharply peaked curves in (a), (b)) indicating a many-body localized state in the Fock space (a stable quasiparticle), while spinful fermions have large support sizes (broadly peaked curves in (c)), indicating a fully delocalized state in the Fock space (a decaying quasiparticle). At high doping density $p>0.33,$ spinful fermions have a small support size (sharply peaked curve in (c)) while magnons and JW fermions have large support sizes (broadly peaked curves in (a), (b)). At the critical value of doping, $p=0.33,$ all three excitations have large widths indicating that they are bad quasiparticles. Localization and delocalization are further confirmed by a finite size scaling analysis (d-f) for the support sizes of (d) magnon and (e) JW fermion and (f) spinful fermionic hole quasiparticle states for different doping densities, as a function of the many-body Hilbert space dimension $D.$ The fitted solid lines are to the form $\xi = f D^c.$  
The abscissae show values of $1/\log_2 D,$ while the ordinates show $\log_2 \xi/\log_2 D.$
An intercept $c\approx 1$ signifies a fully many-body delocalized state (bad quasiparticle) while a small value of the intercept $c\approx 0$ corresponds to a many-body localized or fractal (stable quasiparticle) state.
Note that there is an abrupt jump in exponent $c$ in all three cases upon crossing the localization threshold. For all values of  doping except $p=0.33$, where none of the quasiparticles are stable, at all other doping one or more quasiparticles are found to be good.}
\end{figure*}

The support size $\xi_i$ of the quasiparticle state $|\chi_i \prime\rangle$ in the many-body Hilbert space of exact eigenstates is given by $\xi_i=1/P_i,$ where $P_i$ is the inverse participation ratio (IPR),
\begin{equation}
	P_i=\sum_{j=1}^{D}|a_{ij}|^4.
\end{equation}
 The stability of the magnon and the two fermionic quasiparticles is quantitatively determined by performing a finite size scaling analysis of their respective support sizes.
This requires estimation of large number of excited states for which the recently developed FEAST \cite{FEAST} 
eigensolver is used. 

Fig.~\ref{fig:overlap}(a) shows plots of the disorder averaged \footnote{For every disorder realization $\alpha$, the overlaps $|a^{(\alpha)}_{ij}|^2$ are computed, and the combined data for all realizations is binned in equally spaced energy intervals. The plots show the disorder averaged values, $|a_{ij}|^2 = (1/N_{\text{dis}})\sum_{\alpha}|a^{(\alpha)}_{ij}|^2$ vs average many-body energy in the bin.} squares of the overlap, $|a_{ij}|^2$, for the second lowest  magnon state $|\chi_i\prime\rangle$ with exact many-body states $|\psi_j\rangle$ of the random $t-J$ model as a function of the many-body excitation energy measured from the disorder averaged ground state energy. Curves for different values of hole doping corresponding to under (spin glass phase), over (Fermi liquid phase)  and the putative critical doping $p=p_c=1/3$ are shown. Figs.~1(b) and 1(c) respectively show similar plots for respectively, the second lowest  Jordan-Wigner (JW) fermion and a spinful fermionic (hole) quasiparticle. At low doping densities $p<p_c,$ the magnon and JW quasiparticles have relatively small energy widths when compared to spinful quasiparticles. At high doping densities $p>p_c,$ the magnon and JW quasiparticles have large widths compared to the spinful quasiparticles. We also note that JW quasiparticles are generally less stable than magnons but still good quasiparticles in the underdoped phase. The plots shown are disorder averaged over $\sim 10^3$ configurations. 
Low-lying states have been chosen for stability analysis here as they are more stable than the higher energy ones and also easier to compute.
The above overlap plots include the effects of both disorder and interaction related broadening, with the former not associated with decay.

In the SI, we present plots for lifetime broadening $\Delta \epsilon$ as a function of quasiparticle energy $\epsilon,$ which is analogous to $\text{Im}\Sigma$ studied elsewhere \cite{georges2013bad,cha2020linear}. The expected superlinear energy dependence of $\Delta \epsilon$ for low-lying Landau quasiparticles in the overdoped regime is clearly seen. In the underdoped regime, we see a flat and large $\Delta \epsilon$ expected for bad Landau quasiparticles. For magnons, the overlaps of many-body states with the quasiparticle wavefunction essentially determine the dynamical susceptibility \cite{shackleton2021quantum}. Starting from the underdoped side, we note the broadening of magnon quasiparticles with doping as well as energy, qualitatively consistent with Ref \cite{shackleton2021quantum}. However our decay width analysis does not give a clear signature of the quasiparticle stability and transition, for which we find the IPR method to be better.

 To place our above observations on a more quantitative footing, and also determine the stabilities near critical doping, we perform a finite size scaling analysis (for details see \cite{aman_kitaev,altshuler1997}) of the quasiparticle support sizes in the many-body Hilbert space. 

 If the support size $\xi$ of a quasiparticle state scales with the dimensionality $D$ of the Fock space as $\xi \sim D,$ the quasiparticle is said to be fully many-body delocalized state, and has a finite lifetime that is inversely proportional to its energy width extracted from the support size. If on the other hand, the support size scales as $\xi \sim D^0,$ the quasiparticle state is many-body localized in the Fock space - such a state has an infinite lifetime. A third possibility may also occurs where the scaling of the support size is fractal, $\xi \sim D^c$, with $c<1$ - this represents a nonergodic delocalized state, which is also a good quasiparticle for our purposes.
A perturbative treatment in Ref. \cite{rivas2002numerical} shows that quasiparticle broadening calculated from imaginary part of self-energy ($\text{Im}\Sigma$) is equivalent to an IPR-based analysis.
 
Figure \ref{fig:overlap} (d-f) shows a plot of $(\log_2(\xi))/(\log_2(D))$ versus  $1/\log_2(D)$ for (d) magnon, (e) JW fermions  and (f) spinful fermionic hole quasiparticles. The fits are to the law  $\xi\sim f D^c,$ where $f\equiv f(D)$ has an arbitrary but weaker than power-law dependence on $D.$ Since the interaction term generally does not connect all fermions or magnon quasiparticle states to every many body state even in the fully delocalized phase, we expect $f<1$ in the delocalized regime. In the many-body localized phase, although $c=0,$ the support size $\xi \geq 1,$ and in general also an increasing function of $D;$ consequently, $f(D)\geq 1$ is expected to monotonously increase with $D.$ Fitted lines with positive (negative) slope $(-\log_2(1/f))$ thus correspond to localized (delocalized) phases. For small doping densities ($p<p_c$), the magnon state (Fig.~1(d)) has a small support size, a positive slope $(-\log_2(1/f))$ and $c\approx 0$ corresponding to magnon localization and stability, while beyond the critical doping $p_c=0.33,$ support size of the magnon scales exponentially with $c\approx 1,$ and has negative slope corresponding to full delocalization and instability. Elsewhere in the underdoped phase $p<p_c,$ we observe for the magnon a negative slope with intercept $c<1$ corresponding to a nonergodic (fractal) delocalized state, still corresponding to well-defined quasiparticles. The same behavior is shown by the JW fermions (Fig.~1(e)) although they are systematically more unstable than the magnons in all doping regimes. The quasiparticle stability gets reversed for the spinful fermions (Fig.~1(f)), which are stable for $p>p_c$ and unstable in the underdoped regime. These scaling trends are consistent with the behavior seen in the plots (a-c) of the overlap squares. 
\begin{figure}
	\includegraphics[width=\columnwidth]{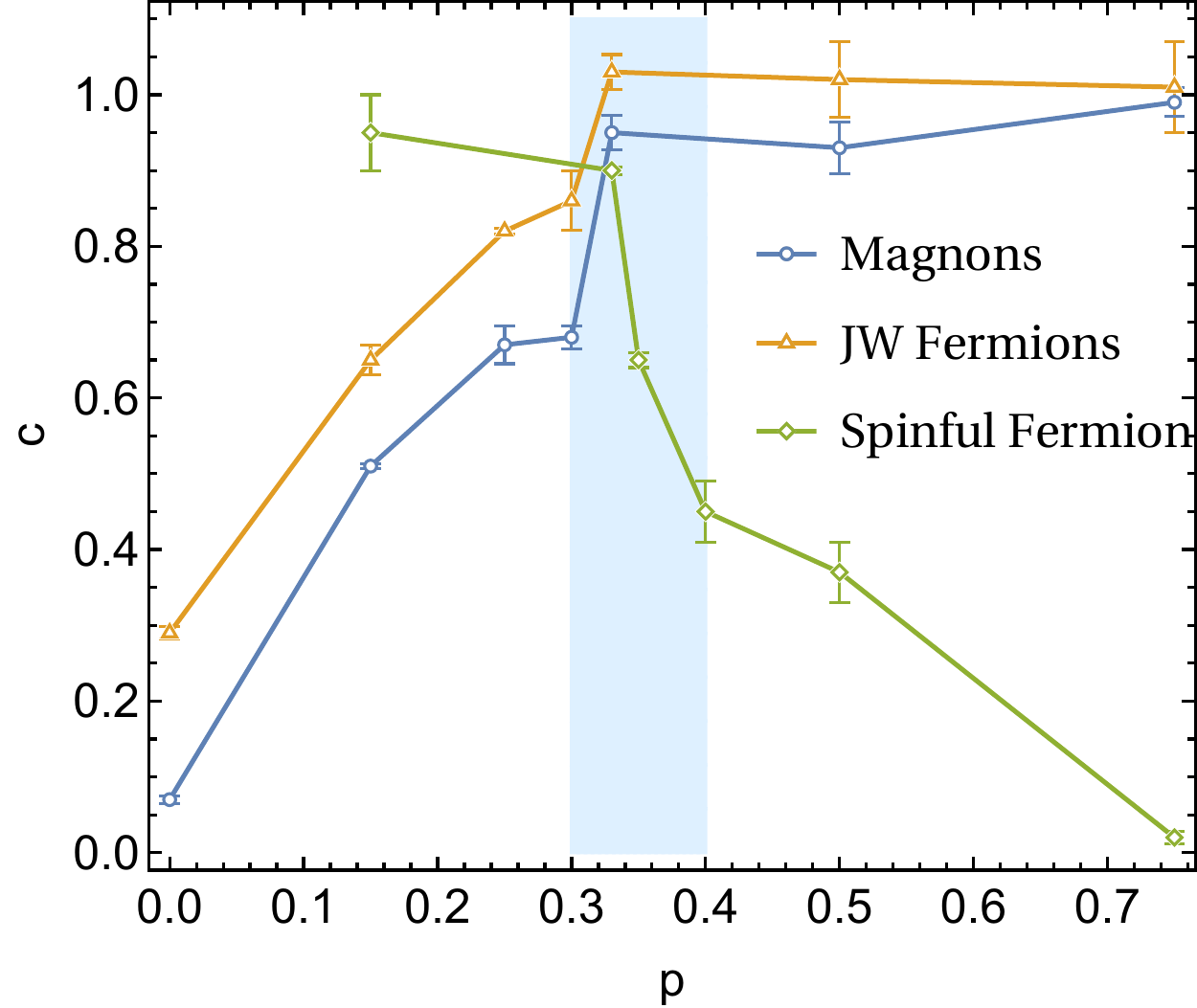}
	\caption{\label{fig:intercept}Plot shows the MBL scaling exponent $c$ versus doping density for all three quasiparticles. A sharp jump in $c$ (to values $c\approx 1$) is seen for all quasiparticles upon crossing to the doping regime where they are unstable.}
\end{figure}
We now discuss the behavior in the vicinity of $p_c.$ Note that the dimensionality $D$ of the Fock space takes its maximum value at $p_c=1/3;$ in the thermodynamic limit it is easily checked that at this critical doping, $D\sim 3^N.$ This indicates three degrees of freedom at each site, signifying some emergent symmetry at this point. In comparison, at zero doping, there are only two degrees of freedom at each site ($D=2^N$), and at very high values of the doping, $D$ increases slower than exponential.
From the scaling analysis in Fig. \ref{fig:overlap} (d-f), it is clear that the exponent $c\approx 1$ for all three types of excitations when $p\approx p_c = 0.33,$ consistent with the understanding from earlier field-theoretical studies\cite{joshi2020deconfined} that there are no good quasiparticles at critical doping in this model. 

Figure \ref{fig:intercept} shows the plot of the intercept $c$ versus doping density $p.$ A sharp step in $c$ is observed near the vicinity of $p_c$ for all quasiparticles. The change in $c$ near $p_c$ for JW fermions is evident (showing they are good quasiparticles that decay in the overdoped side), but less sharp than the jump for magnons, which happen to be better quasiparticles. Interestingly, similar jumps in the scaling exponent have been seen in MBL transitions in disordered XXZ Heisenberg chains \cite{rivas2002numerical}.


In conclusion, using a many-body localization treatment, we confirm the existence of two phases as a function of doping density - a spin glass phase at low doping and a Fermi liquid phase at high values of doping. At large doping density $p>0.33,$ spinful fermionic excitations are well-defined while at small doping $p<0.33,$ magnons and emergent Jordan-Wigner fermions are well-defined. Magnons and spinful fermions of comparable energy are not observed to be simultaneously good quasiparticles for any value of the doping.  Near the critical point $p=0.33,$ we find that all these three excitations are bad quasiparticles. 
A central observation is that the deconfined critical point is associated with a localization transition in the many-body Hilbert space. The existence of stable Jordan-Wigner fermions in the spin glass phase could be of experimental significance. In the ground state of the underdoped (spin glass) phase, the JW quasiparticle density is fairly high, at $(1-p)/2.$ Our analysis lends support to the recent view that these quasiparticles are likely responsible via their coupling to phonons for the anomalously large thermal Hall conductivity seen in the underdoped cuprates (concomitant with poor electrical conductivity).   
\begin{acknowledgments}
     AK and VT acknowledge support of the Department of Atomic Energy, Government of India, under Project Identification No. RTI 4002, and the Department of Theoretical Physics, TIFR, for computational resources. SS was supported by the U.S. National Science Foundation grant No. DMR-2002850 by the Simons Collaboration on Ultra-Quantum Matter which is a grant from the Simons Foundation (651440, S.S.)
\end{acknowledgments}
\appendix
\section{Case of a good quasiparticle: comparison of single disorder realization and disorder averaging}
\begin{figure}
	\includegraphics[width=\columnwidth]{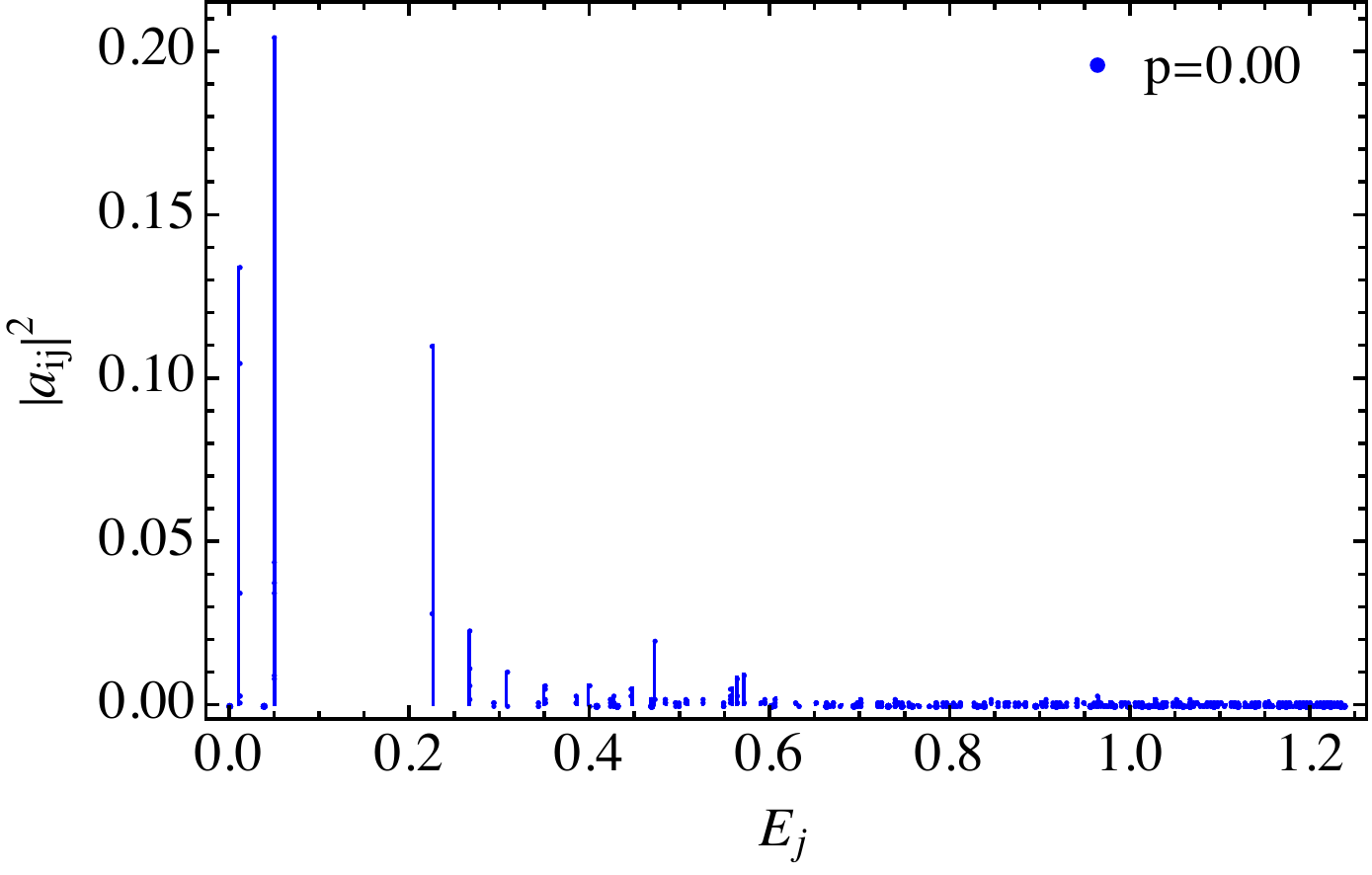}
	\caption{\label{fig:overlap_single} Distribution of the overlap square of the second lowest magnon quasiparticle as function of many-body excitation energy $E_j$ at zero doping, $p=0,$ where the magnon is stable. The plot is for one disorder realization, and the system size is $N=15.$ The quasiparticle has significant overlap with around five ($O(1)$) many-body states. }
\end{figure}

Figure \ref{fig:overlap_single} shows the plot of overlap squares $|a_{ij}|^2$ for the second lowest magnon quasiparticle at zero doping, and in one disorder realization. The magnon quasiparticle state significantly overlaps with only around five (i.e. $O(1)$) exact many body states, even though the Hilbert space has $2^{15}$ states. This proves the quasiparticle is very good. We generate such data for the second lowest magnon for different disorder realizations, and then perform the disorder average. For each disorder realization, the second lowest magnon is similarly found to have significant overlaps with a very small number of many-body states. This is shown in Fig. \ref{fig:IPR} for the distribution of support sizes for 1000 disorder realizations, which is seen to be dominated at support size of around five. 

\begin{figure}
	\includegraphics[width=\columnwidth]{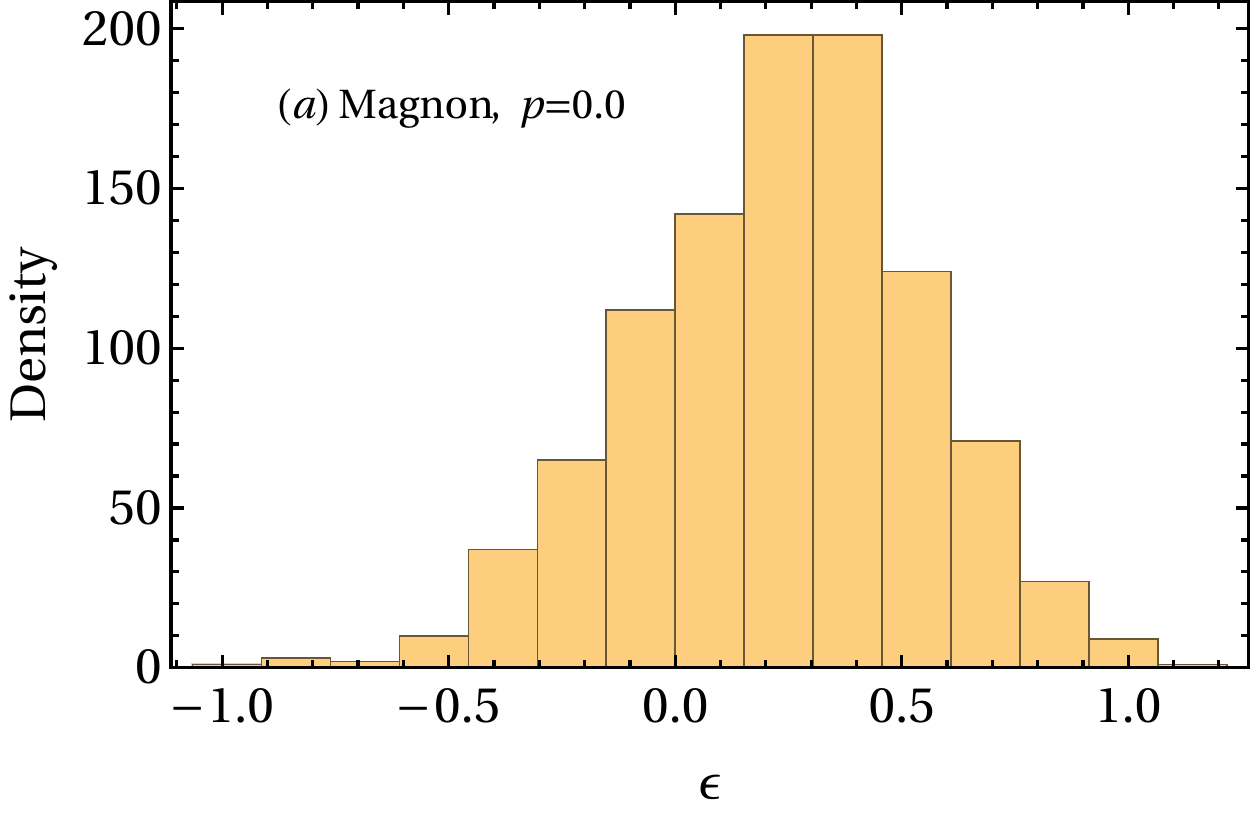}
	\caption{\label{fig:averageE} Plot showing a histogram of the mean excitation energy $\epsilon$ (after quantum averaging ) for a stable (second lowest) magnon at zero doping, 15-site system.  The $\epsilon$ are measured from the disorder averaged ground state energy for 1000 disorder configurations. The distribution peaks at a small nonzero positive energy although there is a large spread on account of the disorder. Note that in a small fraction of samples, the excitation energy is lower than the disorder averaged ground state energy - this is on account of rare disorder configurations. }
\end{figure}

\begin{figure}
	\includegraphics[width=\columnwidth]{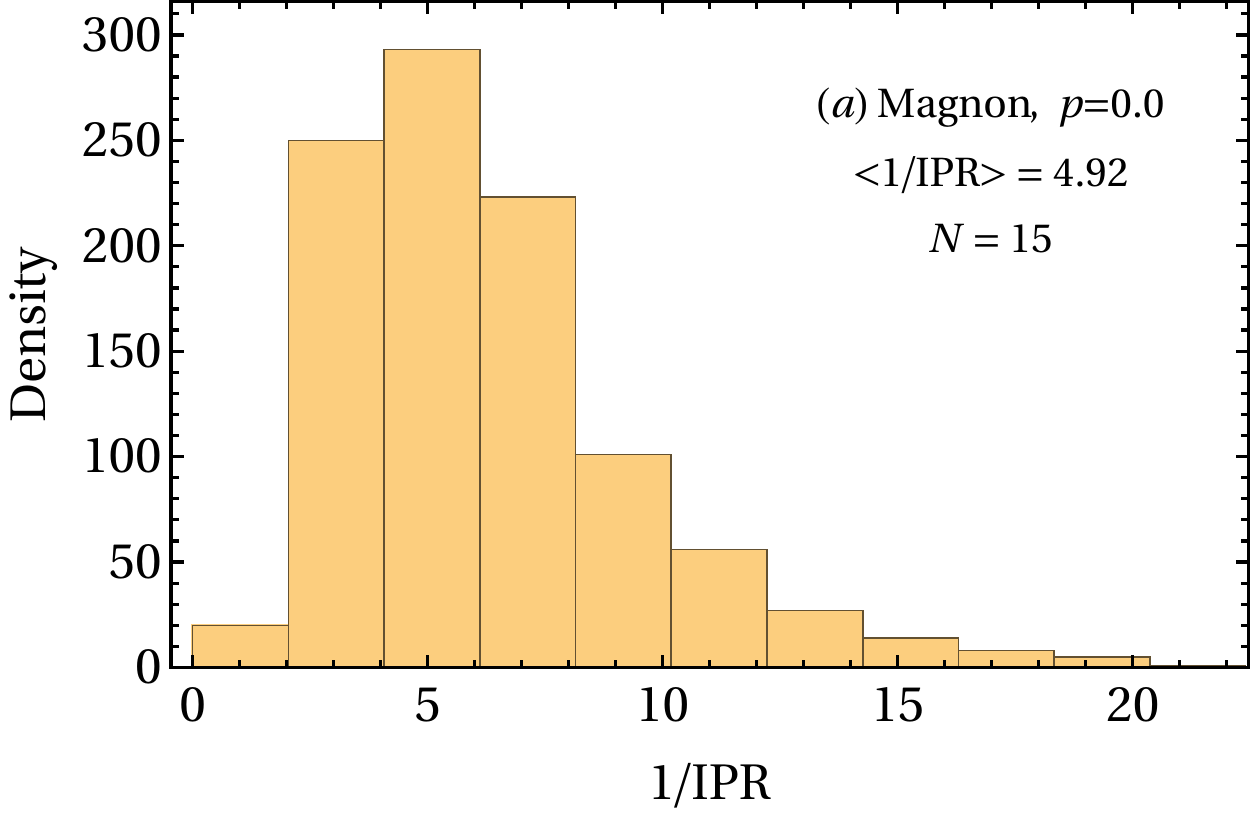}
	\caption{\label{fig:IPR} Plot showing histogram of support sizes ($1/\text{IPR}$) for a stable (second lowest) magnon at zero doping for a 15-site system. The distribution is sharply peaked at a small value $\approx 5$ although the number of states is $2^{15}.$}
\end{figure}

Even though Fig. \ref{fig:overlap_single} and Fig. \ref{fig:IPR} show that this magnon is a very good quasiparticle, we also see from Fig. \ref{fig:averageE} (see also Fig. 1 (left panel) in the manuscript) that the energy spread of this state is not very small, and is comparable to the disorder strength. This spread is not a sign of quasiparticle decay since it's a noninteracting effect arising from different disorder realizations giving different energies for the many-body states.

It is clear from the above example that even when the overlap squares of individual disorder realizations indicate a good quasiparticle, it is not manifest in the disorder averaged plots of the same quantity. Instead, we found that the support size distribution conveys this information more accurately, even after disorder averaging. Physically, this is because a good quasiparticle always has a small support size over a wide rangle of disorder configurations.

\section{Energy width of quasiparticle}
\begin{figure}
	\includegraphics[width=\columnwidth]{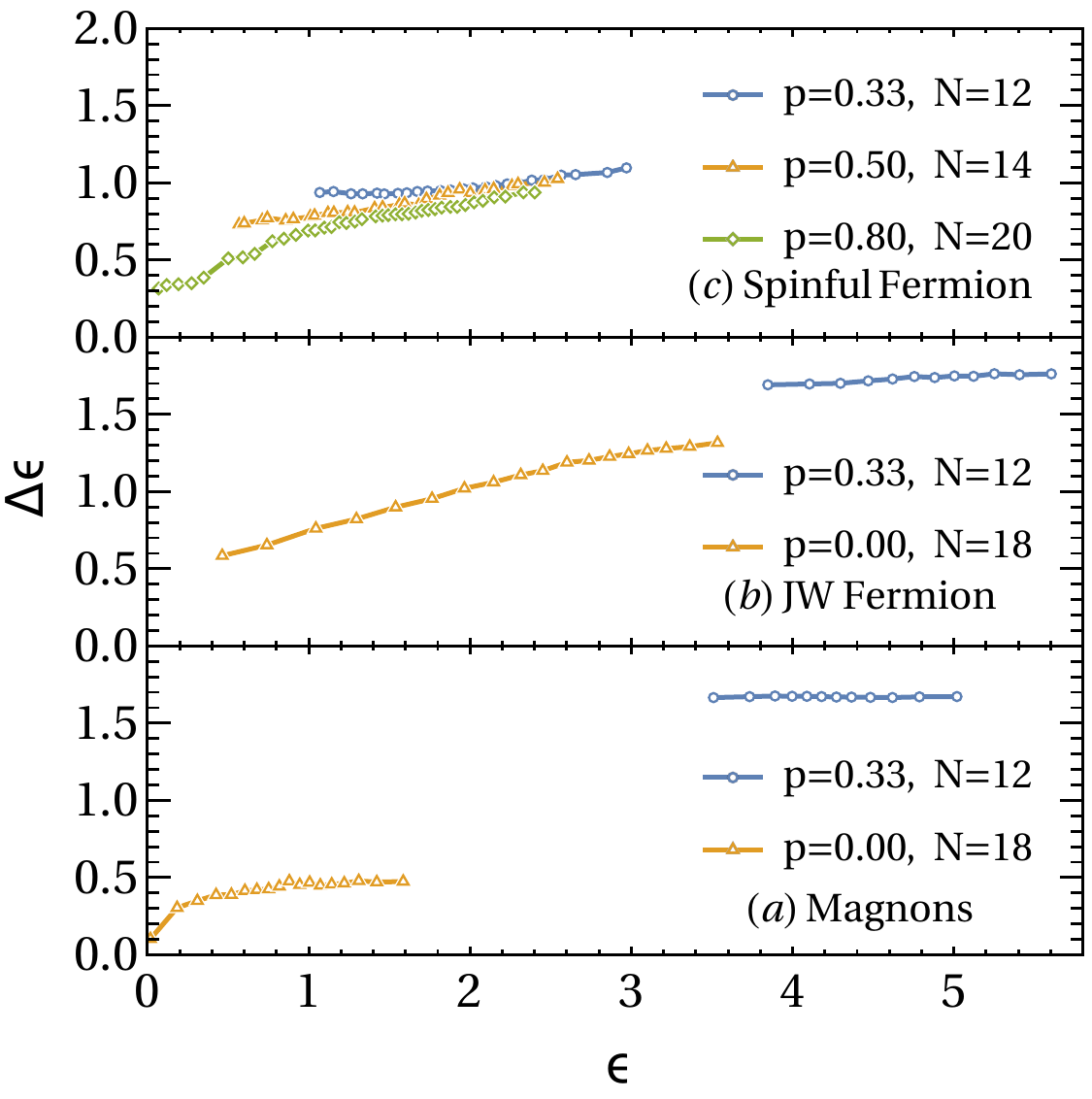}
	\caption{\label{fig:width} Plot shows the evolution of the disorder averaged quasiparticle energy spread $\Delta \epsilon $ versus its mean excitation energy. See Eq. (\ref{eq:spread}) for an explanation. Good quasiparticles are characterized by relatively small quasiparticle widths, and bad ones have large and flat $\Delta \epsilon.$ On the overdoped side, low-energy spinful quasiparticles have a super-linear increase of energy spread, consistent with landau Fermi liquid theory.}
\end{figure}

In the literature, the lifetime broadening (for, say, Landau quasiparticles) in the doped Mott insulators is usually extracted from the imaginary part of the single particle self energy, $\text{Im}\Sigma$ \cite{georges2013bad,cha2020linear}. Although our IPR-based method does not use Green functions, it is possible to obtain a similar measure of the quasiparticle width. For this purpose, we first calculate the energy spread of the quasiparticle for a single disorder realization, and then we perform the average over disorder configurations. For a good quasiparticle, the spread will be small for typical realizations of disorder. For a single disorder realization, $\alpha,$ the energy spread $\Delta \epsilon_{i}^{(\alpha)}$ of the $i^{\text{th}}$ quasiparticle is given by
\begin{align}
(\Delta \epsilon_{i}^{(\alpha)})^2 & = \sum_{j}|a_{ij}^{(\alpha)}|^2 (E_{j}^{\alpha})^2 - \left[\sum_{j}|a_{ij}^{(\alpha)}|^2 E_{j}^{(\alpha)}\right]^2,
\end{align}
where $E_j^{(\alpha)}$ are the many-body energies for this realization. The disorder averaged spread is given by
\begin{align}
(\Delta\epsilon_{i})^2 & = \frac{1}{N_{\text{dis}}}\sum_{\alpha}(\Delta \epsilon_{i}^{(\alpha)})^2,
\label{eq:spread}
\end{align}
where $N_{\text{dis}}$ is the number of disorder realizations (taken to be 1000 in our case).

Fig. \ref{fig:width} shows the disorder averaged energy width $\Delta \epsilon$ versus mean energy $\epsilon$ of the quasiparticle. For the spinful fermions (Landau quasiparticles), far on the overdoped side (here $p=0.80$), $\Delta \epsilon$ starts at a small value and curves upwards, reminiscent of the $\text{Im}\Sigma\sim \epsilon^2$ for a Landau quasiparticle, and at higher energies it flattens out at a certain large value, which is comparable to the quasiparticle bandwidth. At critical doping ($p=0.33$), $\Delta\epsilon$ starts at a large value and varies little with $\epsilon.$ It is also gapped (even though the many-body spectrum is not), indicating it's not a good approximation for a low energy excitation. Closer to the critical doping, but still on the overdoped side (here $p=0.5$), $\Delta \epsilon$ starts at larger values, still appears to increase super-linearly with $\epsilon,$ and soon reaches the saturation value of the order of the quasiparticle bandwidth. A similar trend is seen for magnons and JW fermions. For the magnons at zero doping, $\Delta\epsilon$ starts at small values for low-lying $\epsilon,$ but soon saturates at a relatively small value. The JW fermions on the underdoped side show a sharper increase of $\Delta \epsilon$ similar to that shown by Landau quasiparticles on the overdoped sice. with At critical doping, both magnons and JW fermions have large gaps, and $\Delta\epsilon$ is flat, at a relatively high value comparable to the quasiparticle bandwidth. The trends (e.g. stability and energy dependence of $\Delta\epsilon$ Landau quasiparticles on the overdoped side) are in qualitative agreement with analyses based on other techniques such as DMFT \cite{cha2020linear,georges2013bad}, although quantitative comparison is not possible because our $\Delta\epsilon$ is not exactly $\text{Im}\Sigma$ of a single electron Green function, as the latter is extracted from a part of the spectral function that corresponds to the quasiparticle pole. Note that the transition from stable to unstable quasipaticles does not reveal sharply in our calculations of $\Delta\epsilon$ but is readily detected in the finite size scaling of the IPR.
Note that 

\begin{thebibliography}{26}
\expandafter\ifx\csname natexlab\endcsname\relax\def\natexlab#1{#1}\fi
\expandafter\ifx\csname bibnamefont\endcsname\relax
  \def\bibnamefont#1{#1}\fi
\expandafter\ifx\csname bibfnamefont\endcsname\relax
  \def\bibfnamefont#1{#1}\fi
\expandafter\ifx\csname citenamefont\endcsname\relax
  \def\citenamefont#1{#1}\fi
\expandafter\ifx\csname url\endcsname\relax
  \def\url#1{\texttt{#1}}\fi
\expandafter\ifx\csname urlprefix\endcsname\relax\def\urlprefix{URL }\fi
\providecommand{\bibinfo}[2]{#2}
\providecommand{\eprint}[2][]{\url{#2}}

\bibitem[{\citenamefont{{Frachet} et~al.}(2020)\citenamefont{{Frachet},
  {Vinograd}, {Zhou}, {Benhabib}, {Wu}, {Mayaffre}, {Kr{\"a}mer},
  {Ramakrishna}, {Reyes}, {Debray} et~al.}}]{frachet2020hidden}
\bibinfo{author}{\bibfnamefont{M.}~\bibnamefont{{Frachet}}},
  \bibinfo{author}{\bibfnamefont{I.}~\bibnamefont{{Vinograd}}},
  \bibinfo{author}{\bibfnamefont{R.}~\bibnamefont{{Zhou}}},
  \bibinfo{author}{\bibfnamefont{S.}~\bibnamefont{{Benhabib}}},
  \bibinfo{author}{\bibfnamefont{S.}~\bibnamefont{{Wu}}},
  \bibinfo{author}{\bibfnamefont{H.}~\bibnamefont{{Mayaffre}}},
  \bibinfo{author}{\bibfnamefont{S.}~\bibnamefont{{Kr{\"a}mer}}},
  \bibinfo{author}{\bibfnamefont{S.~K.} \bibnamefont{{Ramakrishna}}},
  \bibinfo{author}{\bibfnamefont{A.~P.} \bibnamefont{{Reyes}}},
  \bibinfo{author}{\bibfnamefont{J.}~\bibnamefont{{Debray}}},
  \bibnamefont{et~al.}, \bibinfo{journal}{Nature Physics}
  \textbf{\bibinfo{volume}{16}}, \bibinfo{pages}{1064} (\bibinfo{year}{2020}),
  \eprint{1909.10258}.

\bibitem[{\citenamefont{Grissonnanche et~al.}(2019)\citenamefont{Grissonnanche,
  Legros, Badoux, Lefran{\c{c}}ois, Zatko, Lizaire, Lalibert{\'e}, Gourgout,
  Zhou, Pyon et~al.}}]{grissonnanche2019giant}
\bibinfo{author}{\bibfnamefont{G.}~\bibnamefont{Grissonnanche}},
  \bibinfo{author}{\bibfnamefont{A.}~\bibnamefont{Legros}},
  \bibinfo{author}{\bibfnamefont{S.}~\bibnamefont{Badoux}},
  \bibinfo{author}{\bibfnamefont{E.}~\bibnamefont{Lefran{\c{c}}ois}},
  \bibinfo{author}{\bibfnamefont{V.}~\bibnamefont{Zatko}},
  \bibinfo{author}{\bibfnamefont{M.}~\bibnamefont{Lizaire}},
  \bibinfo{author}{\bibfnamefont{F.}~\bibnamefont{Lalibert{\'e}}},
  \bibinfo{author}{\bibfnamefont{A.}~\bibnamefont{Gourgout}},
  \bibinfo{author}{\bibfnamefont{J.-S.} \bibnamefont{Zhou}},
  \bibinfo{author}{\bibfnamefont{S.}~\bibnamefont{Pyon}}, \bibnamefont{et~al.},
  \bibinfo{journal}{Nature} \textbf{\bibinfo{volume}{571}},
  \bibinfo{pages}{376} (\bibinfo{year}{2019}).

\bibitem[{\citenamefont{{Arrachea} and
  {Rozenberg}}(2002)}]{arrachea2002infinite}
\bibinfo{author}{\bibfnamefont{L.}~\bibnamefont{{Arrachea}}} \bibnamefont{and}
  \bibinfo{author}{\bibfnamefont{M.~J.} \bibnamefont{{Rozenberg}}},
  \bibinfo{journal}{\prb} \textbf{\bibinfo{volume}{65}}, \bibinfo{eid}{224430}
  (\bibinfo{year}{2002}), \eprint{cond-mat/0203537}.

\bibitem[{\citenamefont{Grissonnanche et~al.}(2020)\citenamefont{Grissonnanche,
  Th{\'e}riault, Gourgout, Boulanger, Lefran{\c{c}}ois, Ataei, Lalibert{\'e},
  Dion, Zhou, Pyon et~al.}}]{grissonnanche2020chiral}
\bibinfo{author}{\bibfnamefont{G.}~\bibnamefont{Grissonnanche}},
  \bibinfo{author}{\bibfnamefont{S.}~\bibnamefont{Th{\'e}riault}},
  \bibinfo{author}{\bibfnamefont{A.}~\bibnamefont{Gourgout}},
  \bibinfo{author}{\bibfnamefont{M.-E.} \bibnamefont{Boulanger}},
  \bibinfo{author}{\bibfnamefont{E.}~\bibnamefont{Lefran{\c{c}}ois}},
  \bibinfo{author}{\bibfnamefont{A.}~\bibnamefont{Ataei}},
  \bibinfo{author}{\bibfnamefont{F.}~\bibnamefont{Lalibert{\'e}}},
  \bibinfo{author}{\bibfnamefont{M.}~\bibnamefont{Dion}},
  \bibinfo{author}{\bibfnamefont{J.-S.} \bibnamefont{Zhou}},
  \bibinfo{author}{\bibfnamefont{S.}~\bibnamefont{Pyon}}, \bibnamefont{et~al.},
  \bibinfo{journal}{Nature Physics} \textbf{\bibinfo{volume}{16}},
  \bibinfo{pages}{1108} (\bibinfo{year}{2020}).

\bibitem[{\citenamefont{Samajdar et~al.}(2019)\citenamefont{Samajdar, Scheurer,
  Chatterjee, Guo, Xu, and Sachdev}}]{samajdar2019enhanced}
\bibinfo{author}{\bibfnamefont{R.}~\bibnamefont{Samajdar}},
  \bibinfo{author}{\bibfnamefont{M.~S.} \bibnamefont{Scheurer}},
  \bibinfo{author}{\bibfnamefont{S.}~\bibnamefont{Chatterjee}},
  \bibinfo{author}{\bibfnamefont{H.}~\bibnamefont{Guo}},
  \bibinfo{author}{\bibfnamefont{C.}~\bibnamefont{Xu}}, \bibnamefont{and}
  \bibinfo{author}{\bibfnamefont{S.}~\bibnamefont{Sachdev}},
  \bibinfo{journal}{Nature Phys.} \textbf{\bibinfo{volume}{15}},
  \bibinfo{pages}{1290} (\bibinfo{year}{2019}), \eprint{1903.01992}.

\bibitem[{\citenamefont{{Dalla Piazza} et~al.}(2015)\citenamefont{{Dalla
  Piazza}, {Mourigal}, {Christensen}, {Nilsen}, {Tregenna-Piggott}, {Perring},
  {Enderle}, {McMorrow}, {Ivanov}, and {R{\o}nnow}}}]{dalla2015fractional}
\bibinfo{author}{\bibfnamefont{B.}~\bibnamefont{{Dalla Piazza}}},
  \bibinfo{author}{\bibfnamefont{M.}~\bibnamefont{{Mourigal}}},
  \bibinfo{author}{\bibfnamefont{N.~B.} \bibnamefont{{Christensen}}},
  \bibinfo{author}{\bibfnamefont{G.~J.} \bibnamefont{{Nilsen}}},
  \bibinfo{author}{\bibfnamefont{P.}~\bibnamefont{{Tregenna-Piggott}}},
  \bibinfo{author}{\bibfnamefont{T.~G.} \bibnamefont{{Perring}}},
  \bibinfo{author}{\bibfnamefont{M.}~\bibnamefont{{Enderle}}},
  \bibinfo{author}{\bibfnamefont{D.~F.} \bibnamefont{{McMorrow}}},
  \bibinfo{author}{\bibfnamefont{D.~A.} \bibnamefont{{Ivanov}}},
  \bibnamefont{and} \bibinfo{author}{\bibfnamefont{H.~M.}
  \bibnamefont{{R{\o}nnow}}}, \bibinfo{journal}{Nature Physics}
  \textbf{\bibinfo{volume}{11}}, \bibinfo{pages}{62} (\bibinfo{year}{2015}),
  \eprint{1501.01767}.

\bibitem[{\citenamefont{Kitaev}(2006)}]{kitaev2006anyons}
\bibinfo{author}{\bibfnamefont{A.}~\bibnamefont{Kitaev}},
  \bibinfo{journal}{Annals of Physics} \textbf{\bibinfo{volume}{321}},
  \bibinfo{pages}{2} (\bibinfo{year}{2006}).

\bibitem[{\citenamefont{Yokoi et~al.}(2021)\citenamefont{Yokoi, Ma, Kasahara,
  Kasahara, Shibauchi, Kurita, Tanaka, Nasu, Motome, Hickey
  et~al.}}]{yokoi2021half}
\bibinfo{author}{\bibfnamefont{T.}~\bibnamefont{Yokoi}},
  \bibinfo{author}{\bibfnamefont{S.}~\bibnamefont{Ma}},
  \bibinfo{author}{\bibfnamefont{Y.}~\bibnamefont{Kasahara}},
  \bibinfo{author}{\bibfnamefont{S.}~\bibnamefont{Kasahara}},
  \bibinfo{author}{\bibfnamefont{T.}~\bibnamefont{Shibauchi}},
  \bibinfo{author}{\bibfnamefont{N.}~\bibnamefont{Kurita}},
  \bibinfo{author}{\bibfnamefont{H.}~\bibnamefont{Tanaka}},
  \bibinfo{author}{\bibfnamefont{J.}~\bibnamefont{Nasu}},
  \bibinfo{author}{\bibfnamefont{Y.}~\bibnamefont{Motome}},
  \bibinfo{author}{\bibfnamefont{C.}~\bibnamefont{Hickey}},
  \bibnamefont{et~al.}, \bibinfo{journal}{Science}
  \textbf{\bibinfo{volume}{373}}, \bibinfo{pages}{568} (\bibinfo{year}{2021}).

\bibitem[{\citenamefont{Kasahara et~al.}(2018)\citenamefont{Kasahara, Ohnishi,
  Mizukami, Tanaka, Ma, Sugii, Kurita, Tanaka, Nasu, Motome
  et~al.}}]{kasahara2018majorana}
\bibinfo{author}{\bibfnamefont{Y.}~\bibnamefont{Kasahara}},
  \bibinfo{author}{\bibfnamefont{T.}~\bibnamefont{Ohnishi}},
  \bibinfo{author}{\bibfnamefont{Y.}~\bibnamefont{Mizukami}},
  \bibinfo{author}{\bibfnamefont{O.}~\bibnamefont{Tanaka}},
  \bibinfo{author}{\bibfnamefont{S.}~\bibnamefont{Ma}},
  \bibinfo{author}{\bibfnamefont{K.}~\bibnamefont{Sugii}},
  \bibinfo{author}{\bibfnamefont{N.}~\bibnamefont{Kurita}},
  \bibinfo{author}{\bibfnamefont{H.}~\bibnamefont{Tanaka}},
  \bibinfo{author}{\bibfnamefont{J.}~\bibnamefont{Nasu}},
  \bibinfo{author}{\bibfnamefont{Y.}~\bibnamefont{Motome}},
  \bibnamefont{et~al.}, \bibinfo{journal}{Nature}
  \textbf{\bibinfo{volume}{559}}, \bibinfo{pages}{227} (\bibinfo{year}{2018}).

\bibitem[{\citenamefont{Varma}(2020)}]{varma2020colloquium}
\bibinfo{author}{\bibfnamefont{C.~M.} \bibnamefont{Varma}},
  \bibinfo{journal}{Rev. Mod. Phys.} \textbf{\bibinfo{volume}{92}},
  \bibinfo{pages}{031001} (\bibinfo{year}{2020}),
  \urlprefix\url{https://link.aps.org/doi/10.1103/RevModPhys.92.031001}.

\bibitem[{\citenamefont{Cha et~al.}(2020)\citenamefont{Cha, Wentzell,
  Parcollet, Georges, and Kim}}]{cha2020linear}
\bibinfo{author}{\bibfnamefont{P.}~\bibnamefont{Cha}},
  \bibinfo{author}{\bibfnamefont{N.}~\bibnamefont{Wentzell}},
  \bibinfo{author}{\bibfnamefont{O.}~\bibnamefont{Parcollet}},
  \bibinfo{author}{\bibfnamefont{A.}~\bibnamefont{Georges}}, \bibnamefont{and}
  \bibinfo{author}{\bibfnamefont{E.-A.} \bibnamefont{Kim}},
  \bibinfo{journal}{Proceedings of the National Academy of Sciences}
  \textbf{\bibinfo{volume}{117}}, \bibinfo{pages}{18341}
  (\bibinfo{year}{2020}), ISSN \bibinfo{issn}{1091-6490}, \eprint{2002.07181},
  \urlprefix\url{http://dx.doi.org/10.1073/pnas.2003179117}.

\bibitem[{\citenamefont{Guo et~al.}(2020)\citenamefont{Guo, Gu, and
  Sachdev}}]{guo2020linear}
\bibinfo{author}{\bibfnamefont{H.}~\bibnamefont{Guo}},
  \bibinfo{author}{\bibfnamefont{Y.}~\bibnamefont{Gu}}, \bibnamefont{and}
  \bibinfo{author}{\bibfnamefont{S.}~\bibnamefont{Sachdev}},
  \bibinfo{journal}{Annals Phys.} \textbf{\bibinfo{volume}{418}},
  \bibinfo{pages}{168202} (\bibinfo{year}{2020}), \eprint{2004.05182}.

\bibitem[{\citenamefont{{Sachdev} and {Ye}}(1993)}]{SY}
\bibinfo{author}{\bibfnamefont{S.}~\bibnamefont{{Sachdev}}} \bibnamefont{and}
  \bibinfo{author}{\bibfnamefont{J.}~\bibnamefont{{Ye}}},
  \bibinfo{journal}{Phys. Rev. Lett.} \textbf{\bibinfo{volume}{70}},
  \bibinfo{pages}{3339} (\bibinfo{year}{1993}), \eprint{cond-mat/9212030}.

\bibitem[{\citenamefont{Kitaev}(2015)}]{kitaev_talk}
\bibinfo{author}{\bibfnamefont{A.}~\bibnamefont{Kitaev}}, \bibinfo{journal}{USA
  April 2015}  (\bibinfo{year}{2015}).

\bibitem[{\citenamefont{{Chowdhury} et~al.}(2021)\citenamefont{{Chowdhury},
  {Georges}, {Parcollet}, and {Sachdev}}}]{tJreview}
\bibinfo{author}{\bibfnamefont{D.}~\bibnamefont{{Chowdhury}}},
  \bibinfo{author}{\bibfnamefont{A.}~\bibnamefont{{Georges}}},
  \bibinfo{author}{\bibfnamefont{O.}~\bibnamefont{{Parcollet}}},
  \bibnamefont{and} \bibinfo{author}{\bibfnamefont{S.}~\bibnamefont{{Sachdev}}}
  (\bibinfo{year}{2021}), \eprint{2109.05037}.

\bibitem[{\citenamefont{Joshi et~al.}(2020)\citenamefont{Joshi, Li,
  Tarnopolsky, Georges, and Sachdev}}]{joshi2020deconfined}
\bibinfo{author}{\bibfnamefont{D.~G.} \bibnamefont{Joshi}},
  \bibinfo{author}{\bibfnamefont{C.}~\bibnamefont{Li}},
  \bibinfo{author}{\bibfnamefont{G.}~\bibnamefont{Tarnopolsky}},
  \bibinfo{author}{\bibfnamefont{A.}~\bibnamefont{Georges}}, \bibnamefont{and}
  \bibinfo{author}{\bibfnamefont{S.}~\bibnamefont{Sachdev}},
  \bibinfo{journal}{Phys. Rev. X} \textbf{\bibinfo{volume}{10}},
  \bibinfo{pages}{021033} (\bibinfo{year}{2020}), \eprint{1912.08822}.

\bibitem[{\citenamefont{Tarnopolsky et~al.}(2020)\citenamefont{Tarnopolsky, Li,
  Joshi, and Sachdev}}]{tarnopolsky2020metal}
\bibinfo{author}{\bibfnamefont{G.}~\bibnamefont{Tarnopolsky}},
  \bibinfo{author}{\bibfnamefont{C.}~\bibnamefont{Li}},
  \bibinfo{author}{\bibfnamefont{D.~G.} \bibnamefont{Joshi}}, \bibnamefont{and}
  \bibinfo{author}{\bibfnamefont{S.}~\bibnamefont{Sachdev}},
  \bibinfo{journal}{Phys. Rev. B} \textbf{\bibinfo{volume}{101}},
  \bibinfo{pages}{205106} (\bibinfo{year}{2020}), \eprint{2002.12381}.

\bibitem[{\citenamefont{Altshuler et~al.}(1997)\citenamefont{Altshuler, Gefen,
  Kamenev, and Levitov}}]{altshuler1997}
\bibinfo{author}{\bibfnamefont{B.~L.} \bibnamefont{Altshuler}},
  \bibinfo{author}{\bibfnamefont{Y.}~\bibnamefont{Gefen}},
  \bibinfo{author}{\bibfnamefont{A.}~\bibnamefont{Kamenev}}, \bibnamefont{and}
  \bibinfo{author}{\bibfnamefont{L.~S.} \bibnamefont{Levitov}},
  \bibinfo{journal}{Phys. Rev. Lett.} \textbf{\bibinfo{volume}{78}},
  \bibinfo{pages}{2803} (\bibinfo{year}{1997}),
  \urlprefix\url{https://link.aps.org/doi/10.1103/PhysRevLett.78.2803}.

\bibitem[{\citenamefont{Kumar and Tripathi}(2020)}]{aman_kitaev}
\bibinfo{author}{\bibfnamefont{A.}~\bibnamefont{Kumar}} \bibnamefont{and}
  \bibinfo{author}{\bibfnamefont{V.}~\bibnamefont{Tripathi}},
  \bibinfo{journal}{Phys. Rev. B} \textbf{\bibinfo{volume}{102}},
  \bibinfo{pages}{100401} (\bibinfo{year}{2020}),
  \urlprefix\url{https://link.aps.org/doi/10.1103/PhysRevB.102.100401}.

\bibitem[{\citenamefont{Polizzi}(2009)}]{FEAST}
\bibinfo{author}{\bibfnamefont{E.}~\bibnamefont{Polizzi}},
  \bibinfo{journal}{Phys. Rev. B} \textbf{\bibinfo{volume}{79}},
  \bibinfo{pages}{115112} (\bibinfo{year}{2009}),
  \urlprefix\url{https://link.aps.org/doi/10.1103/PhysRevB.79.115112}.

\bibitem[{\citenamefont{{Shackleton} et~al.}(2021)\citenamefont{{Shackleton},
  {Wietek}, {Georges}, and {Sachdev}}}]{shackleton2021quantum}
\bibinfo{author}{\bibfnamefont{H.}~\bibnamefont{{Shackleton}}},
  \bibinfo{author}{\bibfnamefont{A.}~\bibnamefont{{Wietek}}},
  \bibinfo{author}{\bibfnamefont{A.}~\bibnamefont{{Georges}}},
  \bibnamefont{and}
  \bibinfo{author}{\bibfnamefont{S.}~\bibnamefont{{Sachdev}}},
  \bibinfo{journal}{Phys. Rev. Lett.} \textbf{\bibinfo{volume}{126}},
  \bibinfo{eid}{136602} (\bibinfo{year}{2021}), \eprint{2012.06589}.

\bibitem[{\citenamefont{{Dumitrescu} et~al.}(2021)\citenamefont{{Dumitrescu},
  {Wentzell}, {Georges}, and {Parcollet}}}]{Dumi21}
\bibinfo{author}{\bibfnamefont{P.~T.} \bibnamefont{{Dumitrescu}}},
  \bibinfo{author}{\bibfnamefont{N.}~\bibnamefont{{Wentzell}}},
  \bibinfo{author}{\bibfnamefont{A.}~\bibnamefont{{Georges}}},
  \bibnamefont{and}
  \bibinfo{author}{\bibfnamefont{O.}~\bibnamefont{{Parcollet}}}
  (\bibinfo{year}{2021}), \eprint{2103.08607}.

\bibitem[{\citenamefont{{Banerjee} et~al.}(2016)\citenamefont{{Banerjee},
  {Bridges}, {Yan}, {Aczel}, {Li}, {Stone}, {Granroth}, {Lumsden}, {Yiu},
  {Knolle} et~al.}}]{banerjee2016proximate}
\bibinfo{author}{\bibfnamefont{A.}~\bibnamefont{{Banerjee}}},
  \bibinfo{author}{\bibfnamefont{C.~A.} \bibnamefont{{Bridges}}},
  \bibinfo{author}{\bibfnamefont{J.~Q.} \bibnamefont{{Yan}}},
  \bibinfo{author}{\bibfnamefont{A.~A.} \bibnamefont{{Aczel}}},
  \bibinfo{author}{\bibfnamefont{L.}~\bibnamefont{{Li}}},
  \bibinfo{author}{\bibfnamefont{M.~B.} \bibnamefont{{Stone}}},
  \bibinfo{author}{\bibfnamefont{G.~E.} \bibnamefont{{Granroth}}},
  \bibinfo{author}{\bibfnamefont{M.~D.} \bibnamefont{{Lumsden}}},
  \bibinfo{author}{\bibfnamefont{Y.}~\bibnamefont{{Yiu}}},
  \bibinfo{author}{\bibfnamefont{J.}~\bibnamefont{{Knolle}}},
  \bibnamefont{et~al.}, \bibinfo{journal}{Nature Materials}
  \textbf{\bibinfo{volume}{15}}, \bibinfo{pages}{733} (\bibinfo{year}{2016}),
  \eprint{1504.08037}.

\bibitem[{\citenamefont{{Grusdt} et~al.}(2019)\citenamefont{{Grusdt}, {Bohrdt},
  and {Demler}}}]{grusdt2019microscopic}
\bibinfo{author}{\bibfnamefont{F.}~\bibnamefont{{Grusdt}}},
  \bibinfo{author}{\bibfnamefont{A.}~\bibnamefont{{Bohrdt}}}, \bibnamefont{and}
  \bibinfo{author}{\bibfnamefont{E.}~\bibnamefont{{Demler}}},
  \bibinfo{journal}{\prb} \textbf{\bibinfo{volume}{99}}, \bibinfo{eid}{224422}
  (\bibinfo{year}{2019}), \eprint{1901.01113}.

\bibitem[{\citenamefont{Deng et~al.}(2013)\citenamefont{Deng, Mravlje, Ferrero,
  Kotliar, Georges et~al.}}]{georges2013bad}
\bibinfo{author}{\bibfnamefont{X.}~\bibnamefont{Deng}},
  \bibinfo{author}{\bibfnamefont{J.}~\bibnamefont{Mravlje}},
  \bibinfo{author}{\bibfnamefont{M.}~\bibnamefont{Ferrero}},
  \bibinfo{author}{\bibfnamefont{G.}~\bibnamefont{Kotliar}},
  \bibinfo{author}{\bibfnamefont{A.}~\bibnamefont{Georges}},
  \bibnamefont{et~al.}, \bibinfo{journal}{Physical review letters}
  \textbf{\bibinfo{volume}{110}}, \bibinfo{pages}{086401}
  (\bibinfo{year}{2013}).

\bibitem[{\citenamefont{Rivas et~al.}(2002)\citenamefont{Rivas, Mucciolo, and
  Kamenev}}]{rivas2002numerical}
\bibinfo{author}{\bibfnamefont{A.~M.} \bibnamefont{Rivas}},
  \bibinfo{author}{\bibfnamefont{E.~R.} \bibnamefont{Mucciolo}},
  \bibnamefont{and} \bibinfo{author}{\bibfnamefont{A.}~\bibnamefont{Kamenev}},
  \bibinfo{journal}{Physical Review B} \textbf{\bibinfo{volume}{65}},
  \bibinfo{pages}{155309} (\bibinfo{year}{2002}).

\end{thebibliography}

\end{document}